\documentclass[11pt,twoside]{article}
\usepackage{graphicx,epsfig,multirow,amsmath,bm,natbib}
\usepackage{CS18}

\newcommand{\Rsun}[1]{#1\,\mathrm{R} \,\!\scriptscriptstyle \sun\!}
\newcommand{\Osun}[1]{#1\,\Omega \,\!\scriptscriptstyle \sun\!}
\newcommand{\Prt}{\mathrm{Pr}}
\newcommand{\Prm}{\mathrm{Pm}}

\newcommand{\orho}{\overline{\rho}}
\newcommand{\ddr}[1]{\frac{\partial #1}{\partial r}}
\newcommand{\ddtime}[1]{\frac{\partial #1}{\partial t}}
\newcommand{\sint}{\sin{\theta}}
\newcommand{\vv}{\mathbf{v}}
\newcommand{\vB}{\mathbf{B}}

\newcommand{\vcr}{\mathrm{v_r}}
\newcommand{\avg}[1]{\langle #1 \rangle}
\newcommand{\bigavg}[1]{\Big\langle #1 \Big\rangle}
\newcommand{\pht}{\hat{\boldsymbol{\phi}}}
\newcommand{\grad}{\boldsymbol{\nabla}}

\newcommand{\curl}{\boldsymbol{\nabla}\boldsymbol{\times}}
\newcommand{\cnabla}{\boldsymbol{\cdot}\boldsymbol{\nabla}}
\newcommand{\cross}{\boldsymbol{\times}}

\newcommand{\Bct}{B_{\theta}}
\newcommand{\Bcp}{B_{\phi}}

\newcommand{\avgO}{\langle\Omega\rangle}
\newcommand{\avgBp}{\langle B_{\phi} \rangle}

\newcommand{\TME}{\mathrm{T_{ME}}}
\newcommand{\PME}{\mathrm{P_{ME}}}

\newcommand{\TMS}{\mathrm{T_{MS}}}
\newcommand{\PMS}{\mathrm{P_{MS}}}

\newcommand{\TMA}{\mathrm{T_{MA}}}
\newcommand{\PMA}{\mathrm{P_{MA}}}

\newcommand{\TFS}{\mathrm{T_{FS}}}
\newcommand{\PFS}{\mathrm{P_{FS}}}

\newcommand{\TFA}{\mathrm{T_{FA}}}
\newcommand{\PFA}{\mathrm{P_{FA}}}

\newcommand{\TCC}{\mathrm{T_{CC}}}
\newcommand{\PCC}{\mathrm{P_{CC}}}

\newcommand{\TRD}{\mathrm{T_{RD}}}
\newcommand{\PRD}{\mathrm{P_{RD}}}

\newcommand{\MBF}{\langle \vB\rangle}
\newcommand{\MVF}{\langle \vv\rangle}

\newcommand{\MTF}{\langle \Bcp\rangle}
\newcommand{\MPF}{\langle \mathbf{B}_P \rangle}

\newcommand{\mpers}[1]{#1\,m\,s^{-1}}

\begin{document}

\title{Convective Dynamo Simulation with a Grand Minimum}

\author{Kyle Augustson$^1$ with Sacha Brun$^2$, Mark Miesch$^1$, and Juri Toomre$^3$}

\affil{$^1$High Altitude Observatory, 3080 Center Green Drive, Boulder, Colorado USA 80301}
\affil{$^2$Laboratoire AIM Paris-Saclay, CEA/Irfu Universit\'{e} Paris-Diderot CNRS/INSU, 91191 Gif-sur-Yvette}
\affil{$^3$JILA, University of Colorado, Boulder, Colorado USA 80309}

\begin{abstract}
  The global-scale dynamo action achieved in a simulation of a Sun-like star rotating at thrice the
  solar rate is assessed. The 3-D MHD Anelastic Spherical Harmonic (ASH) code, augmented with a
  viscosity minimization scheme, is employed to capture convection and dynamo processes in this
  G-type star. The simulation is carried out in a spherical shell that encompasses 3.8 density scale
  heights of the solar convection zone. It is found that dynamo action with a high degree of time
  variation occurs, with many periodic polarity reversals occurring roughly every 6.2~years. The
  magnetic energy also rises and falls with a regular period. The magnetic energy cycles arise from
  a Lorentz-force feedback on the differential rotation, whereas the processes leading to polarity
  reversals are more complex, appearing to arise from the interaction of convection with the mean
  toroidal fields. Moreover, an equatorial migration of toroidal field is found, which is linked to
  the changing differential rotation, and potentially to a nonlinear dynamo wave. This simulation
  also enters a grand minimum lasting roughly 20~years, after which the dynamo recovers its regular
  polarity cycles.
\end{abstract}

\section{Introduction}

The Sun exhibits many time scales from the ten minute lifetimes of granules to multi-millennial
magnetic activity modulations. One of the most prominent of these scales is the roughly 11-year
sunspot cycle, during which the number of magnetically active regions waxes and wanes. Observations
of the magnetic field at the solar surface reveal complex, hierarchical structures existing on a
vast range of spatial scales. Despite these complexities, large-scale organized spatial patterns
such as Maunder's butterfly diagram, Joy's law, and Hale's polarity law suggest the existence of a
structured large-scale magnetic field within the solar convection zone. In particular, on the Sun's
surface active regions initially emerge at mid-latitudes and appear at increasingly lower latitudes
as the cycle progresses, thus exhibiting equatorward migration \citep[e.g.,][]{hathaway10}.

Such large-scale magnetic phenomenon that vary with the solar cycle are likely being sustained
through convective dynamo action occurring within the solar interior. It has been suspected for at
least 60 years that the crucial ingredients for the solar dynamo are the shear of the differential
rotation and the helical nature of the small-scale convective flows present in the solar convection
zone \citep[e.g.,][]{parker55, steenbeck69, parker77}. Yet even with the advancement to fully
nonlinear global-scale 3-D MHD simulations \citep[e.g.,][]{gilman83,glatzmaier85,brun04,browning06},
achieving dynamo action that exhibits the basic properties of Sun's magnetism has been quite
challenging.  Nonetheless, recent global-scale simulations of convective dynamos have begun to make
substantial contact with some of the properties of the solar dynamo through a wide variety of
numerical methods \citep[e.g.,][]{brown11a,racine11, kapyla12, nelson13a}.

\section{Computational Methods} \label{sec:d3smethods}

The simulation of convection and dynamo action presented here uses the ASH code to evolve the
anelastic equations for a conductive calorically perfect gas in a rotating spherical shell. ASH
solves these equations with a spherical harmonic decomposition of the entropy, magnetic field,
pressure, and mass flux in the horizontal directions. Either a Chebyshev polynomial representation
\citep{clune99,miesch00}, or a fourth order non-uniform finite difference in the radial direction,
resolve radial derivatives.  The radial finite difference derivative scheme is used here. The
solenoidality of the mass flux and magnetic vector fields is maintained through the use of a stream
function formalism \citep{brun04}. The Crank-Nicholson implicit time-stepping method advances the
linear terms in the MHD equations including eddy diffusion, pressure gradients, and
buoyancy. Nonlinear terms such as the convective derivatives, Lorentz forces, and dissipative
processes are handled with a second-order Adams-Bashforth time step. The boundary conditions used
are impenetrable on radial boundaries, with a constant radial gradient of the specific entropy there
as well. The magnetic boundary conditions are perfectly conducting at the lower boundary and
extrapolated as a potential field at the upper boundary. Furthermore, the latest version of the ASH
code is employed, where the necessary MPI communication pathways and memory layout have been
overhauled.

A slope-limited diffusion (SLD) mechanism similar to the scheme presented in \citet{rempel09} and
\citet{fan13} was implemented into the reformulated ASH code. The SLD acts locally to achieve a
spatially monotonic solution by limiting the linearly reconstructed slope in each coordinate
direction. The scheme minimizes the steepest gradients through the action of the minmod limiter,
while the rate of diffusion is regulated by the local velocity. The rate of diffusion is further
reduced through a function $\phi$ of the ratio of the cell-edge difference $\delta q_i$ and the
cell-center difference $\Delta q_i$ in a given direction $i$ for the quantity $q$. Similar to
\citet{rempel09}, this function is defined as $\phi_i=(\mathrm{minmod}[\delta q_i,\Delta
  q_i])^8$. This function limits the action of the diffusion to regions with large differences in
the reconstructed solutions at cell-edges. Since SLD is a nonlinear operator, it is computed in
physical space, incurring the cost of smaller time steps due to the convergence of the grid at the
poles. The resulting diffusion fields are projected back into spectral space and added to the
solution through the Adams-Bashforth scheme used for the other nonlinear terms.

\begin{figure}[t!]
   \begin{center}
     \plotone{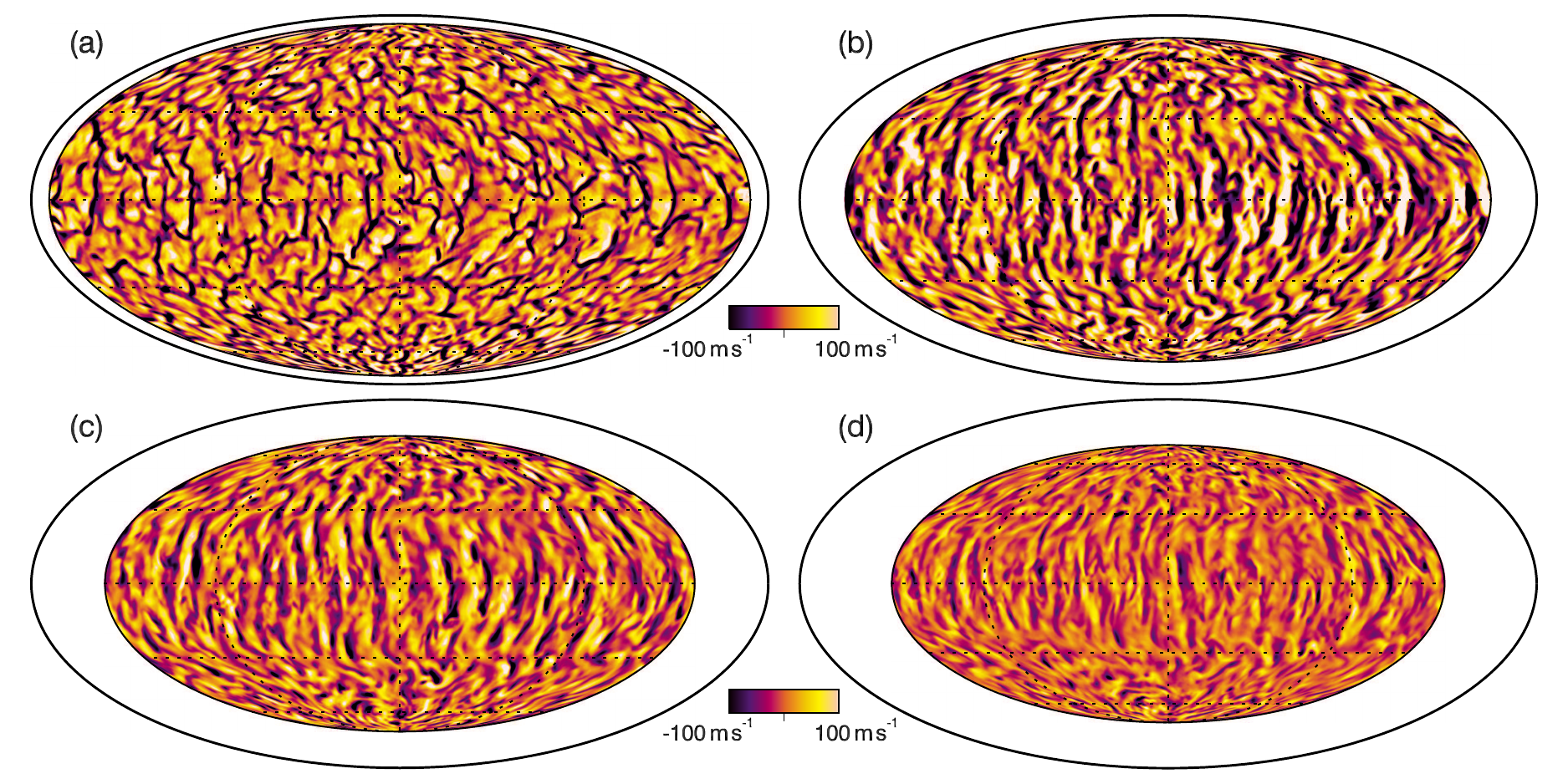} 
     \caption[Radial velocities with depth in a slope-limited ASH simulation]{Snapshots of the
       horizontal convective patterns arising in the radial velocity $v_r$ from the ASH simulation
       presented here that uses slope-limited diffusion are shown in Mollweide projection and
       sampled at four depths: (a) $\Rsun{0.95}$, (b) $\Rsun{0.87}$, (c) $\Rsun{0.80}$, (d)
       $\Rsun{0.75}$. The patterns possess larger-scale columnar convection at low latitudes and the
       smaller-scales at higher latitudes, with downflows dark and upflows in lighter tones. The
       velocity range is the same for each panel, as indicated with the colorbar. \label{fig:d3svr}}
   \end{center}
   \vspace{-0.25truein}
\end{figure}

The simulation encompasses the bulk of the solar convection zone and the full spherical geometry,
though the rotation rate of the frame is $\Omega_0 = 3\Omega_{\sun}$. The domain stretches from the
base of the convection zone at $\Rsun{0.715}$ to $\Rsun{0.965}$, omitting the tachocline and the
deep radiative interior as well as the complex physics of the near-surface layers and approximating
their action with an impenetrable boundary. The SLD has been restricted to act only on the velocity
field in this simulation. This mimics a lower thermal and magnetic Prandtl number ($\Prt$, $\Prm$)
than otherwise obtainable through an elliptic diffusion operator. Yet the entropy and magnetic
fields remain under the influence of an inhomogeneous eddy diffusion, with a radially dependent
entropy diffusion coefficient $\kappa_S$ and magnetic diffusivity $\eta$. The magnitude and form of
these two diffusion coefficients are similar to those of case D3 from \citet{brown10}, with
$\kappa_S , \eta \propto \overline{\rho}^{\; -1/2}$, with $\overline{\rho}$ the spherically
symmetric density. However, this case has about twice the density contrast across the domain, being
45 rather than 26, and has a resolution of $N_r\times N_{\theta} \times N_{\phi} =
200\times256\times512$. In keeping with the ASH nomenclature for the simulations explored in
\citet{brown10}, \citet{brown11a}, and \citet{nelson13a}, this dynamo solution has been called K3S.

\section{Global-Scale Convective Dynamo Action} \label{sec:d3scycles}

Global-scale convective dynamo simulations in rotating spherical shells have recently achieved the
long-sought goal of cyclic magnetic polarity reversals with a multi-decadal period. Moreover, some
of these simulations have illustrated that large-scale dynamo action is possible within the bulk of
the convection zone, even in the absence of a tachocline \citep{brown10}. Global-scale MHD
simulations of a more rapidly rotating Sun with the pseudo-spectral ASH code have produced polarity
reversing dynamo action that possesses strong toroidal wreaths of magnetism that propagate poleward
as a cycle progresses \citep{brown11a,nelson13a}, much like the earlier work of \citet{gilman83} and
\citet{glatzmaier85}. These fields are seated within the convection zone, with much of the magnetic
energy being near the base of the convection zone. The perfectly conducting lower boundary condition
used here and in those simulations requires the field to be horizontal there, which tends to promote
the formation of longitudinal structure in the presence of a differential rotation.

\begin{figure}[t!]
   \begin{center}
     \plotone{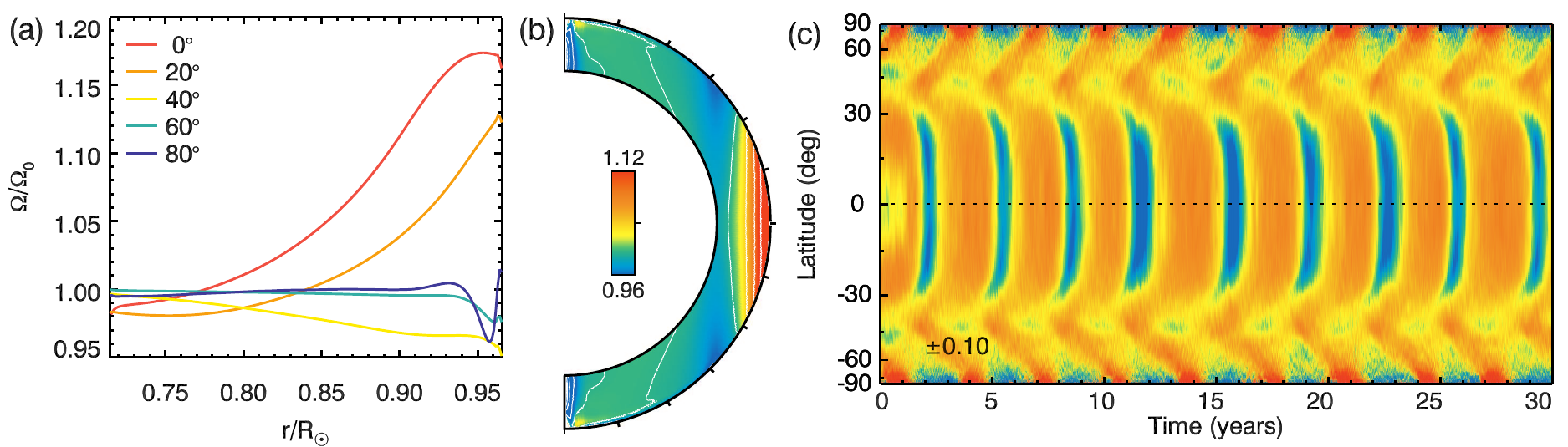} 
     \caption[Properties of the differential rotation in K3S]{The properties of the differential
       rotation in K3S. (a) Cuts at constant latitude through the time-averaged,
       azimuthally-averaged, and normalized angular velocity $\avg{\avgO}/\Omega_0$ (double brackets
       indicating dual averages and $\Omega_0$ the frame rotation rate of $3\Omega_{\sun}$), showing
       a fast equator and slow poles. (b) $\avg{\avgO}/\Omega_0$ shown in the meridional plane,
       illustrating the nearly cylindrical rotation profile.  (c) A time-latitude diagram of
       azimuthally-averaged $\avg{\Delta\Omega}=\avgO-\avg{\avgO}$ in cylindrical projection,
       elucidating the propagation of equatorial and polar branches of a torsional oscillation
       arising from strong Lorentz-force feedback. The color indicates enhanced differential
       rotation in red and periods of slower rotation in blue, with variations of up to $\pm 10$\%
       of the bulk rotation rate. \label{fig:d3somega}}
   \end{center}
   \vspace{-0.25truein}
\end{figure}

Contemporaneously, implicit large-eddy simulations (ILES) have paved the road toward more orderly
long-term cycles in a setting that mimics the solar interior. Indeed, simulations utilizing the
Eulerian-Lagrangian (EULAG) code produce regular polarity cycles occurring roughly every 80 years in
the presence of a tachocline and with the bulk of the magnetic field existing at higher latitudes
\citep{ghizaru10}. This simulation showed radial propagation of structures but little latitudinal
variation during a cycle. More recent simulations of a Sun-like star rotating at $\Osun{3}$ also
produce low-latitude poleward propagating solutions \citep{charbonneau13}. In both cases, such
dynamo action is accomplished through two mechanisms: first by reducing the enthalpy transport of
the largest scales through a simple sub-grid-scale (SGS) model that diminishes thermal perturbations
over a roughly 1.5~year time scale, serving to moderate the global Rossby number; and second through
the ILES formulation of the semi-implicit EULAG MHD code that attempts to maximize the complexity of
the flows and magnetic fields for a given spatial resolution.

Inspired by these recent ASH and EULAG results, an attempt has been made to make contact with both
numerical methods through the incorporation of SLD into ASH with the express goal of achieving a low
effective $\Prt$ and $\Prm$ dynamo. Thus an attempt is made to better mimic the low Prandtl number
solar setting, while keeping the eddy-diffusive approximation for entropy mixing and treating the
reconnection of small-scale magnetic field as diffusive. This effort minimizes the effects of
viscosity, and so extends the inertial range as far as possible for a given resolution. Thus SLD
permits more scales to be captured before entering the dissipation range. With this newly
implemented diffusion minimization scheme, a solution was obtained that possesses four fundamental
features of the solar dynamo: a regular magnetic energy cycle period, and an orderly magnetic
polarity cycle with a period of $\tau_C=6.2$~years, equatorward propagation of magnetic features,
and poleward migration of oppositely signed flux. Furthermore this equilibrium is punctuated by an
interval of relative quiescence, after which the cycle is recovered.

\begin{figure}[t!]
   \begin{center}
     \includegraphics[width=0.5\textwidth]{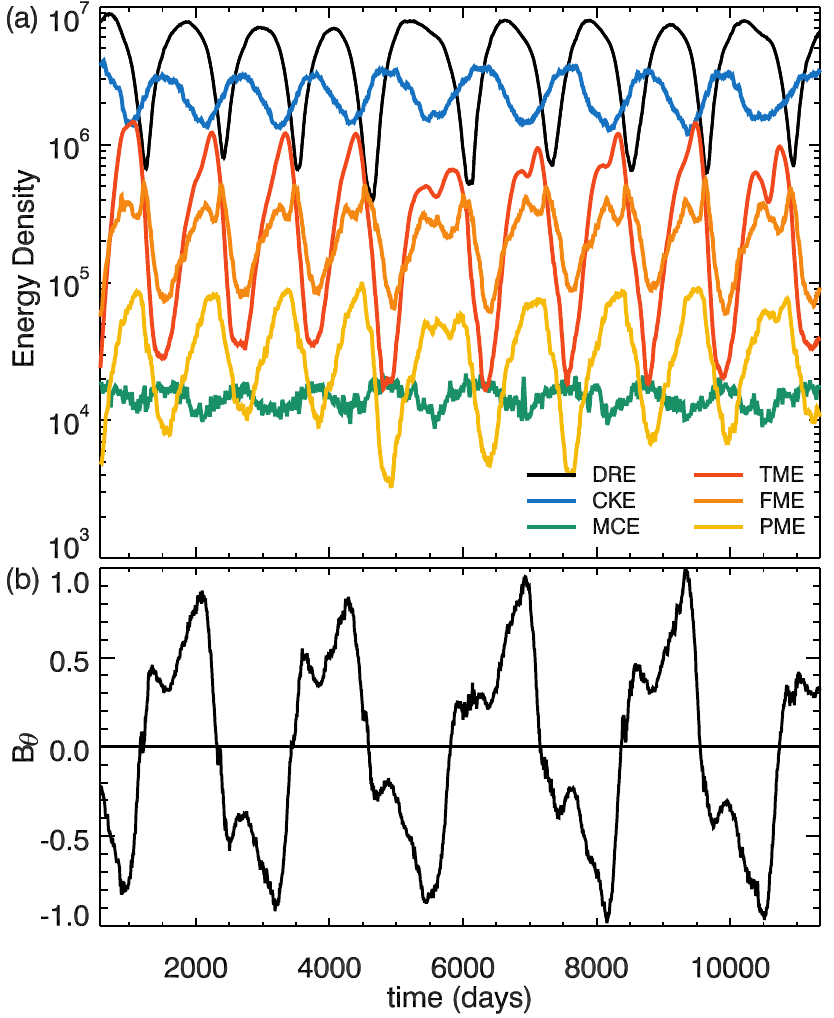}
     \caption[Time variability of $\Bct$ and the volume-averaged energy densities in
       K3S]{Volume-averaged energy densities and latitudinal magnetic field in K3S. (a) Time traces
       of volume-averaged energy densities associated with the differential rotation (DRE),
       turbulent convective kinetic energy (CKE), meridional circulation (MCE), toroidal magnetic
       field (TME), fluctuating magnetic field (FME), and the poloidal magnetic field (PME). The
       magnetic energy cycle is evident. (b) The horizontally-averaged and normalized latitudinal
       magnetic field ($\Bct$) shown at mid-convection zone, illustrating the magnetic polarity
       cycle. \label{fig:d3sscalar}}
   \end{center}
   \vspace{-0.25truein}
\end{figure}

\section{Convective Dynamics in K3S}

Some aspects of the nature of the convective dynamics achieved within K3S are visible in Figure
\ref{fig:d3svr}. In the upper convection zone, the downflows are roughly twice as fast as the
upflows, with the rms upflow being about $\mpers{100}$ and the downflow roughly $\mpers{-200}$. This
decreases with depth, with the asymmetry between the up and downflows nearly vanishing, with the
typical velocity being about $\mpers{10}$. The convection is very columnar reflecting the strength
of the Coriolis force acting upon it, which tends to tilt the convective structures toward the
rotation axis. In particular, when viewed at a constant radius, the convective structures appear as
elongated and north-south aligned flows at low latitudes and smaller scales at higher latitudes (see
Figure \ref{fig:d3svr}). In aggregate, the spatial structure and flow directions along these cells
produce strong Reynolds stresses that act to accelerate the equator and slow the poles. The Reynolds
stresses arising from these turbulent structures are the dominant mechanism that maintains the
differential rotation, as the viscous stresses are quite small in this simulation. The collective
action of these cells also leads to a significant latitudinal enthalpy flux that heats the poles and
sustains the thermal wind balance between the differential rotation and the latitudinal entropy
gradient. The thermal wind in concert with the Reynolds stresses serves to rebuild and maintain the
differential rotation during each cycle. Indeed, a substantial differential rotation is established
and maintained here, as can be seen in Figure \ref{fig:d3somega}. It is also apparent at higher
latitudes and just below the upper boundary that this simulation has a negative radial gradient in
the angular velocity $\Omega$ (Figure \ref{fig:d3somega}a), akin to the near-surface shear layer
that is well-known from helioseismology. The reason for its formation is likely a numerical boundary
layer effect rather than capturing the physical processes present in the solar near-surface shear
layer. However, the Maxwell stresses are certainly important during the latter phases of a magnetic
energy cycle where they act to quench both the convection and the differential rotation, which is
evident in the volume averaged energy traces of Figure \ref{fig:d3sscalar}.

\begin{figure}[t!]
   \begin{center}
   \plotone{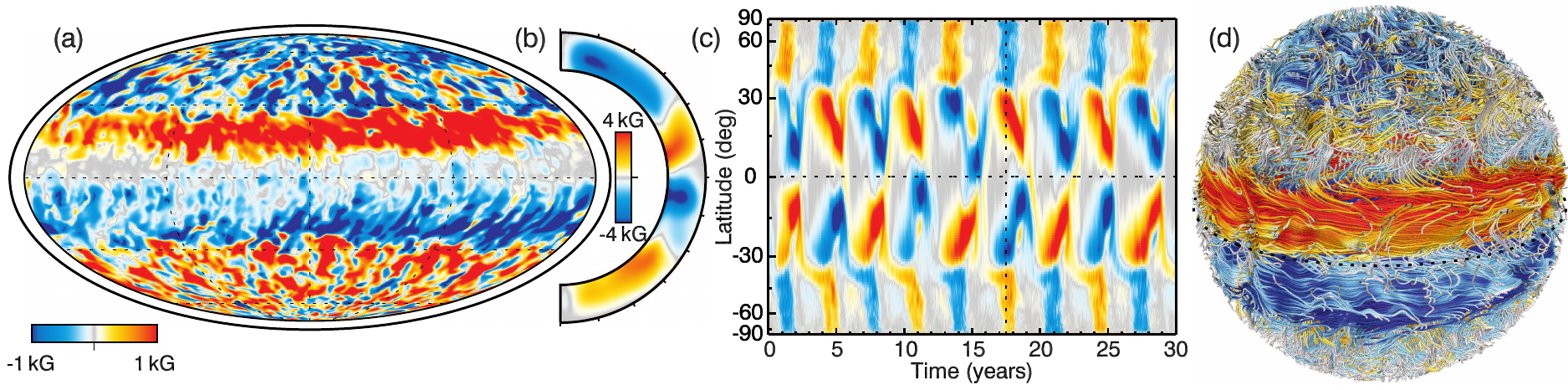}
   \caption[Nature of the toroidal magnetic field in K3S]{Nature of the toroidal magnetic field
     $\Bcp$. (a) Snapshot of the horizontal structure of $\Bcp$ at $\Rsun{0.95}$ shown in Mollweide
     projection, at the time corresponding to the vertical dashed line in (c). This illustrates the
     azimuthal connectivity of the magnetic wreaths, with the polarity of the field such that red
     (blue) tones indicate positive (negative) toroidal field. (b) Azimuthally-averaged $\avgBp$
     also time-averaged over a single energy cycle, depicting the structure of the toroidal field in
     the meridional plane. (c) Time-latitude diagram of $\avgBp$ at $\Rsun{0.95}$ in cylindrical
     projection, exhibiting the equatorward migration of the wreaths from the tangent cylinder at
     roughly $38\deg$ and the poleward propagation of the higher latitude field. The color and
     scaling is as in (a). (d) A rendering of magnetic field lines in the domain colored by the
     magnitude and sign of $\Bcp$, with strong positively oriented field in red, and the strong
     oppositely directed field in blue. \label{fig:d3stfe}}
   \end{center}
   \vspace{-0.25truein}
\end{figure}

The variable nature of the convective patterns over a cycle is an important piece of the K3S dynamo.
Indeed, the magnetic fields disrupt the alignment and correlations of these cells through Lorentz
forces. Particularly, as the field gathers strength during a cycle, the strong azimuthally-connected
toroidal fields tend to create a thermal shadow that weakens the thermal driving of the equatorial
cells. Thus their angular momentum transport is also diminished, which explains why the differential
rotation seen in Figure \ref{fig:d3somega}(b) cannot be fully maintained during the cycle. This is
captured in the modulation of the kinetic energy contained in the fluctuating velocity field, which
here varies by about 50\% throughout the cycle as is visible in Figure \ref{fig:d3sscalar}(a). Such
a mechanism is in keeping with the impacts of strong toroidal fields in the convection zone
suggested by \citet{parker87}. Moreover, strong nonlinear Lorentz force feedbacks have been seen in
other convective dynamo simulations as well \citep{brown11a}, and they have been theoretically
realized for quite some time \citep{malkus75}.

Figure \ref{fig:d3stfe} illustrates the morphology of the toroidal fields in space and time. The
presence of large-scale and azimuthally-connected structures is evident in Figures
\ref{fig:d3stfe}(a, d). Such toroidal structures have been dubbed wreaths \citep{brown10}. In K3S,
there are two counter-polarized, lower-latitude wreaths that form near the point where the tangent
cylinder intersects horizontal surfaces. This site is also where the peak in the latitudinal
gradient of the differential rotation exists for much of a magnetic energy cycle. There are also
polar caps of magnetism of the opposite sense to those at lower latitudes. These caps serve to
moderate the polar differential rotation, which would otherwise tend to accelerate and hence
establish fast polar vortices. The average structure of the wreaths and caps is apparent in Figure
\ref{fig:d3stfe}(b), which is averaged over a single energy cycle (3.1~years). The wreaths appear
rooted at the base of the convection zone, whereas the caps have the bulk of their energy in the
lower convection zone above its base. This is somewhat deceptive as the wreaths are initially
generated higher in the convection zone, while the wreath generation mechanism (primarily the
$\Omega$-effect) migrates equatorward and toward the base of the convection zone over the course of
the cycle. The wreaths obtain their greatest amplitude at the base of the convection zone and thus
appear seated there.

\section{Grand Minima and Long-Term Modulation} \label{sec:d3smin}

\begin{figure}[t!]
   \begin{center}
   \includegraphics[width=0.5\textwidth]{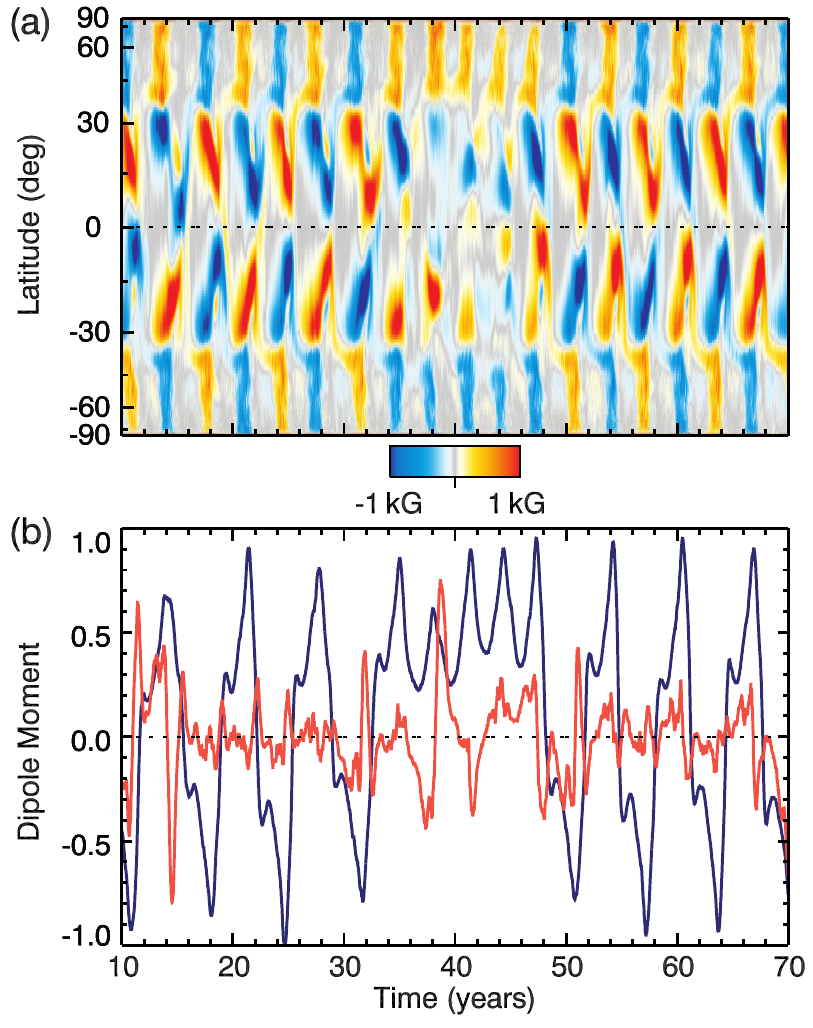}
   \caption[Grand minima in a global-scale convective dynamo simulation]{An interval of magnetic
     quiescence. (a) Time-latitude diagram of $\avgBp$ at $\Rsun{0.95}$ in cylindrical projection,
     picturing the loss and reappearance of cyclic polarity reversals as well as the lower
     amplitude of the wreaths. Strong positive toroidal field is shown as red, negative in blue. (b)
     Normalized magnetic dipole moment (blue) and quadrapole moment normalized by the dipole
     absolute maximum (red). There is a significant rise in the quadrapole moment near reversals,
     suggesting a brief coupling between families of dipolar and quadrapolar dynamo
     modes. \label{fig:d3sint}}
   \end{center}
   \vspace{-0.25truein}
\end{figure}

As with some other dynamo simulations \citep[e.g.,][]{brown11a,augustson13a}, there is also a degree
of long-term modulation of the magnetic cycles in case K3S. Figure \ref{fig:d3sint} shows an
interval of about 20 years during which the polarity cycles are lost, though the magnetic energy
cycles resulting from the nonlinear interaction of the differential rotation and the Lorentz force
remain. During this interval, the magnetic energy in the domain is about 25\% lower, whereas the
energy in the volume encompassed by the lower-latitudes is decreased by 60\%. However, both the
spatial and temporal coherency of the cycles are recovered after this interval and persist for the
last 40~years of the 100~year-long simulation.  Prior to entering this quiescent period, there was
an atypical cycle with only the northern hemisphere exhibiting equatorward propagation. This cycle
also exhibits a prominent loss of the equatorial anti-symmetry in its magnetic polarity. The
subsequent four energy cycles do not reverse their polarity, which is especially evident in the
polar regions, whereas the lower latitudes do seem to attempt such reversals.

Similar quasi-periodic dynamo action has been previously reported on, though only in mean-field
models and in reduced dynamo models and not in fully nonlinear 3D MHD simulations. Some of the most
well-studied quasi-periodic systems utilize reduced dynamo models that couple only a few modes of
the overall system, such as those models of \citet{weiss84}. Weiss uses a simple $\alpha\Omega$
dynamo model. This is accomplished through systems of coupled nonlinear ordinary differential
equations, where the number of included modes is varied. This results in models that range from
fourth to sixth-order in their time derivative. These models admit simple nonlinear oscillatory
solutions. The highest-order model with the largest number of dynamic modes exhibits transitions
from periodic oscillations to chaotic behavior with longer-term modulation akin to the grand minima
in the solar activity cycle.

Later work explored alternative dynamo feedback mechanisms in these low-order systems
\citep{tobias97}. Tobias explored the impacts that modulations in the toroidal magnetic field energy
and that Lorentz-force feedback on the large-scale flows have on the character and evolution of
nonlinear $\alpha\Omega$ dynamo solutions. The former mechanism is shown to affect the parity of the
dynamo, whereas the latter can reduce the efficiency of both the $\alpha$ and $\Omega$ effects in
the limit of low magnetic Prandtl numbers. With an appropriate choice of parameters, some solutions
exhibit a long-term evolution that produces grand extrema, while retaining the basic cycle periods,
as the Sun is observed to do. In particular, it is shown that the magnetic field can be
substantially modulated and effectively remain dipolar. During grand minima the magnetic field is
weak and is no longer a simple dipole. Such behavior is also seen in the simulation here, where its
grand minimum is associated with a much stronger quadrapole moment, as is evident in Figure
\ref{fig:d3sint}b. It also worth noting that the quadrapole component tends to briefly spike during
the polarity reversal, which may be a signal of the reversal mechanism. Indeed, recent work has
explored solar observations over the past three solar cycles to assess the magnetic energy contained
in a large range of spherical harmonic modes \citep{derosa12}. It is seen there that the coupling
between the dipolar and quadrapolar families of modes correlates well with the large-scale polarity
reversals of the Sun. Similar phenomenon seen in the K3S simulation will be explored further in a
forthcoming paper.

\section{Generating Global-Scale Magnetism} \label{sec:d3sgenmag}

The toroidal field shown in Figure \ref{fig:d3stfe} is initially generated and then maintained by
similar processes. During the growth phase of the magnetic field, the shear of the differential
rotation acts to fold and wind the initial poloidal field into toroidal structures. During this
kinematic phase, the mean shear and meridional flows are largely unaffected and can be considered
stationary relative to the time scales of the growing field. However, once the magnetic fields are
strong enough they begin to quench the convective flows that cross them.  The magnetic field
strength becomes saturated as the back reaction of the Lorentz forces increases the alignment of the
velocity field and the magnetic field, which reduces both its generation and at times leads to its
partial destruction.

This configuration of fields establishes a new balance between the various mechanisms of angular
momentum transport needed to sustain the differential rotation when compared to a hydrodynamic
simulation. Here magnetic fields both transport angular momentum as well as block formerly open
channels of transport. For instance, the strong toroidal field serves to reduce the latitudinal
transport of angular momentum by Reynolds stresses, which modifies both the differential rotation
and the meridional circulations within the simulation relative to what would be achieved in a
hydrodynamic simulation.

\begin{figure}[t!]
   \begin{center}
   \plotone{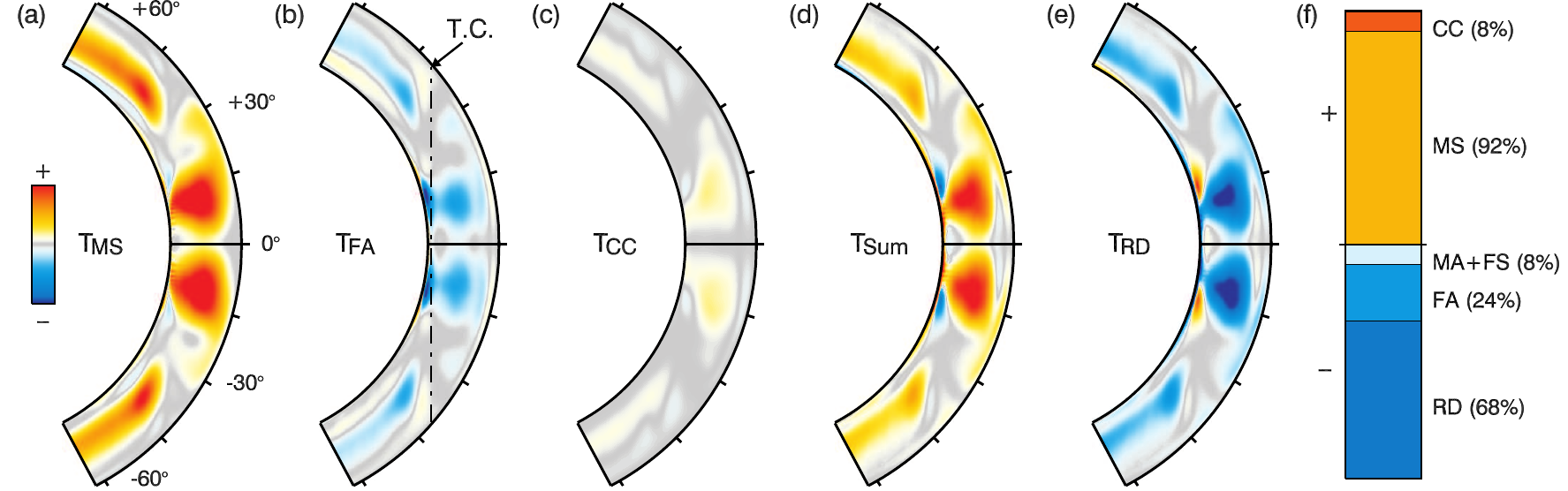}
   \caption[Generation of mean toroidal magnetic energy ($\TME$) in K3S]{Generation of mean toroidal
     magnetic energy ($\TME$) in K3S. The view is from $\pm 60^{\circ}$ to emphasize the equatorial
     regions. Here only the dominant production terms are shown, namely (a) mean shear ($\TMS$), (b)
     fluctuating advection ($\TFA$), with a cylinder tangent to the bottom of the domain (T.C.) is
     indicated (dashed line), (c) compressive correlations ($\TCC$), (d) the sum
     $\TMS+\TFS+\TMA+\TFA+\TCC$, which largely act in concert to balance (e) the resistive diffusion
     of field ($\TRD$). All panels have identical scaling. The mean generation terms shown here
     contribute to the $\TME$ when positive (red), and destroy it if they are negative (blue). (f)
     The relative contribution of each term, with terms in red toned regions adding to the volume
     and time-averaged toroidal energy generation and those in blue dissipating
     energy. \label{fig:d3sphigen}}
   \end{center}
   \vspace{-0.25truein}
\end{figure}

Since the dynamo running within K3S waxes and wanes as time marches forward through many polarity
reversals, the balance between magnetic field generation mechanisms is not instantaneous nor is it
the same at any given time within the cycle. When averaged in time over the full span of the
simulation, though neglecting the interval of the grand minimum, an analysis of the terms
contributing to the generation and destruction of magnetic energy can illustrate those mechanisms
that have the greatest overall influence on the dynamo. The evolving magnetic fields arise from many
competing processes that both produce and destroy magnetic field. These processes are: the shearing
and advection of field, compressive motions, and dissipation through resistive processes. These
production and dissipation terms are most easily understood by an evolution equation for the
magnetic energy contained in the mean magnetic fields. This equation can be broken into its poloidal
and toroidal components $\PME$ and $\TME$ as

\vspace{-0.5truein}
\begin{center}
   \begin{align}
      \displaystyle \ddtime{\TME} &= \frac{\MTF}{8\pi}\bigg[\overbrace{\left.\left(\MBF\cnabla\right)\MVF\right|_{\phi}}^{\TMS}
                    + \overbrace{\bigavg{\left.\left(\vB'\cnabla\right)\vv'}\right|_{\phi}}^{\TFS} 
                    - \overbrace{\left.\left(\MVF\cnabla\right)\MBF\right|_{\phi}}^{\TMA} \nonumber \\
       & - \overbrace{\left.\bigavg{\left(\vv'\cnabla\right)\vB'}\right|_{\phi}}^{\TFA} + \overbrace{\langle\Bcp\vcr\rangle \ddr{\ln{\orho}}}^{\TCC} 
         - \overbrace{\left.\curl\left(\eta\curl\MBF\right)\right|_{\phi}}^{\TRD} \bigg], & \label{eqn:d3smeantme}
   \end{align}
\end{center}

\vspace{-0.5truein}
\begin{center}
   \begin{align}
      \displaystyle \ddtime{\PME} &= \frac{\MPF}{8\pi}\boldsymbol{\cdot}\bigg[\overbrace{\left(\MBF\cnabla\right)\MVF}^{\PMS}
                    + \overbrace{\bigavg{\left(\vB'\cnabla\right)\vv'}}^{\PFS} - \overbrace{\left(\MVF\cnabla\right)\MBF}^{\PMA} 
                    - \overbrace{\bigavg{\left(\vv'\cnabla\right)\vB'}}^{\PFA}  \nonumber \\ 
       &  + \overbrace{\langle\mathbf{B}_P\vcr\rangle \ddr{\ln{\orho}}}^{\PCC} - \overbrace{\curl\left(\eta\curl\MBF\right)}^{\PRD} \bigg], & \label{eqn:d3smeanpme}
   \end{align}
\end{center}

\noindent with the $\langle\rangle$ denoting an average in longitude, $\vv'=\vv-\avg{\vv}$ the
fluctuating velocity, $\vB'=\vB-\avg{\vB}$ the fluctuating magnetic field, and $\avg{\vv}$ and
$\avg{\vB}$ the axisymmetric velocity and magnetic field respectively. A detailed derivation of the
mean-field production terms in spherical coordinates is provided in Appendix A of \citet{brown10}.
Here $\mathrm{T}$ indicates a toroidal magnetic energy generation term and $\mathrm{P}$ a poloidal
energy generation term. The subscripts denote production in turn by mean shear ($\mathrm{MS}$),
fluctuating shear ($\mathrm{FS}$), mean advection ($\mathrm{MA}$), fluctuating advection
($\mathrm{FA}$), compressional correlations ($\mathrm{CC}$), and resistive diffusion
($\mathrm{RD}$).

\begin{figure}[t!]
   \begin{center}
   \plotone{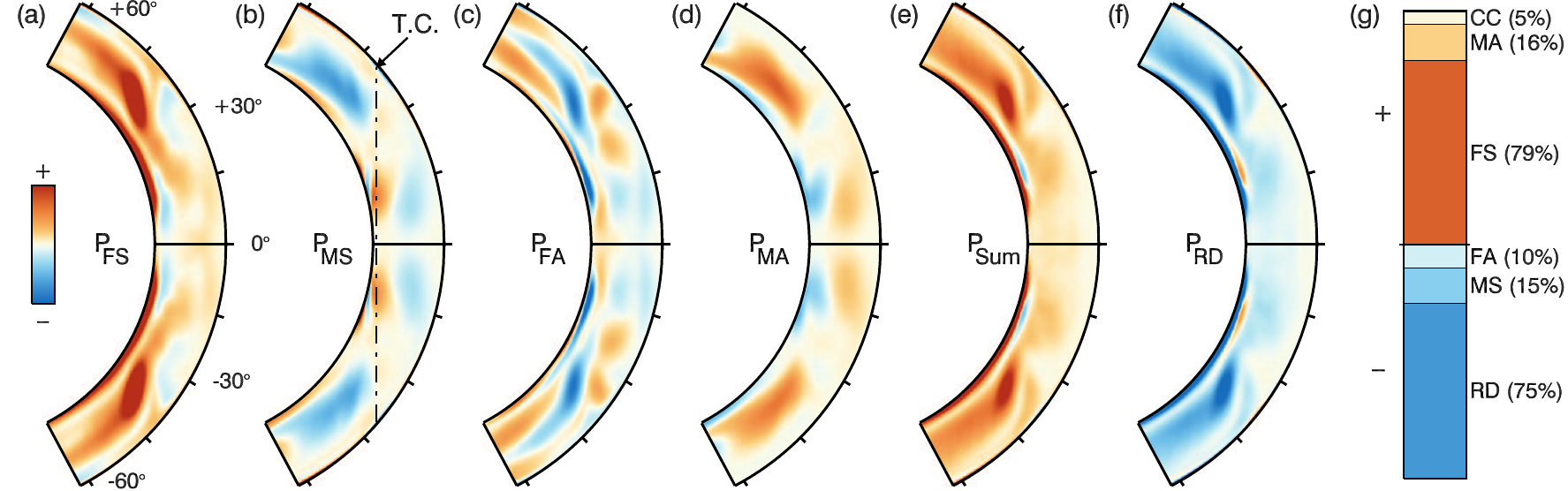}
   \caption[Generation of mean poloidal magnetic energy ($\PME$) in K3S]{Generation of mean poloidal
     magnetic energy ($\PME$) in K3S. The dominant production/destruction terms are shown, namely
     (a) fluctuating shear ($\PFS$), (b) mean shear ($\PMS$), (c) fluctuating advection ($\PFA$),
     (d) mean advection ($\PMA$) and (e) the sum $\PMS+\PFS+\PMA+\PFA+\PCC$, with (f) the resistive
     diffusion of field ($\PRD$).  The mean generation terms shown contribute to the $\PME$ when
     positive (red), and destroy it if negative (blue). (g) The relative contribution of each term,
     with those in red adding to the volume- and time-averaged poloidal energy generation and those
     in blue dissipating energy. \label{fig:d3spolgen}}
   \end{center}
   \vspace{-0.25truein}
\end{figure}

Over a long time average, the time variability of the production of $\TME$ is removed, leaving a
balance between terms that produce and terms that destroy field. Figure \ref{fig:d3sphigen} presents
such a time average of these mean generation terms, involving the entire evolution of case K3S over
an 80~year interval (3300~rotations). In this statistically steady state, the maintenance of the
toroidal wreaths of magnetic field is largely governed by a balance between the production of field
by both the mean shear and the dissipation of field through resistive processes. By comparing
Figures \ref{fig:d3sphigen}(a) and \ref{fig:d3sphigen}(d), it is clear that the mean shear $\TMS$ of
Equation (\ref{eqn:d3smeantme}) is primarily responsible for maintaining the strength of the
wreaths. The production due to the compressibility of the downflows, which can be seen from the form
of $\TCC$, plays a weak but supporting role. In keeping with the dominance of the $\Omega$ effect
($\TMS$ above), the toroidal magnetic energy generation is greatest at latitudes where the
latitudinal gradients in the differential rotation are at their largest. This maximum occurs at
latitudes outside the tangent cylinder in a roughly $15\deg$ swath, though there is also
generation of toroidal field at higher latitudes just inside the tangent cylinder. The resistive
dissipation and fluctuating advection of field act in tandem to destroy toroidal field. Given the
spatial distribution of $\TFA$ in Figure \ref{fig:d3sphigen}b, the action of the fluctuating
advection could be considered as part of an $\alpha$-effect with the small-scale flows tearing at
the large-scale wreaths, transferring energy from the toroidal to the poloidal magnetic fields.

The relative contributions of each term have been measured by integrating them over time and space
$\avg{T_i} = \int T_i dV$, where $i$ is any of the terms in Equation (\ref{eqn:d3smeantme}). This
allows the net action of a given term to be assessed, indicating whether it on average is productive
or destructive of toroidal magnetic energy.  As indicated in Figure \ref{fig:d3sphigen}(f), the bulk
of the production is accomplished by $\TMS$, being about 92\% of the total. The remaining 8\% of the
mean production is due mostly to the compressive motions of the downflows ($\TCC$). In contrast, the
averaged $\TRD$, $\TFA$, $\TMA$, and $\TFS$ processes act in concert to dissipate magnetic energy,
with $\TRD$ responsible for 68\% of the dissipation and $\TFA$ 24\%.

The mechanisms generating the poloidal fields and its associated magnetic energy $\PME$ are given in
Equation (\ref{eqn:d3smeanpme}). In Figure \ref{fig:d3spolgen}, the temporal and longitudinal
averages of the generation terms with the largest average contributions are shown. The average is
again carried out over 80~years. Figure \ref{fig:d3spolgen}(g) shows that in the spatially
integrated sense and unlike in the toroidal energy generation, the fluctuating shear $\PFS$ is the
dominant contributor to the production of poloidal field, generating on average 79\% of the poloidal
energy, while the resistive diffusion $\PRD$ again is the largest dissipator, destroying 75\% of it.
However, other processes contribute to the generation of poloidal field as well. The production of
poloidal field arises through the joint action of the fluctuating shear $\PFS$ and mean advection
$\PMA$, whereas the dissipation of poloidal magnetic energy arises from the combination of resistive
diffusion $\PRD$ and mean shear $\PMS$. The bulk of the production occurs near the lower boundary
and inside the tangent cylinder, being associated with the fluctuating velocities and magnetic
fields of the more isotropic convective cells at higher latitudes. The generation of poloidal field
has two primary balances: the first is between production through the fluctuating shear and
resistive decay; the second is between the mean shear and the mean advection. The overall balances
of magnetic energy generation are subtle given the spatial separation between the primary regions of
toroidal and poloidal field generation, with poloidal field being built near the base of the
convection zone and at higher latitudes and the toroidal field being produced at low latitudes in
the upper convection zone.

\begin{figure}[t!]
   \begin{center}
     \plotone{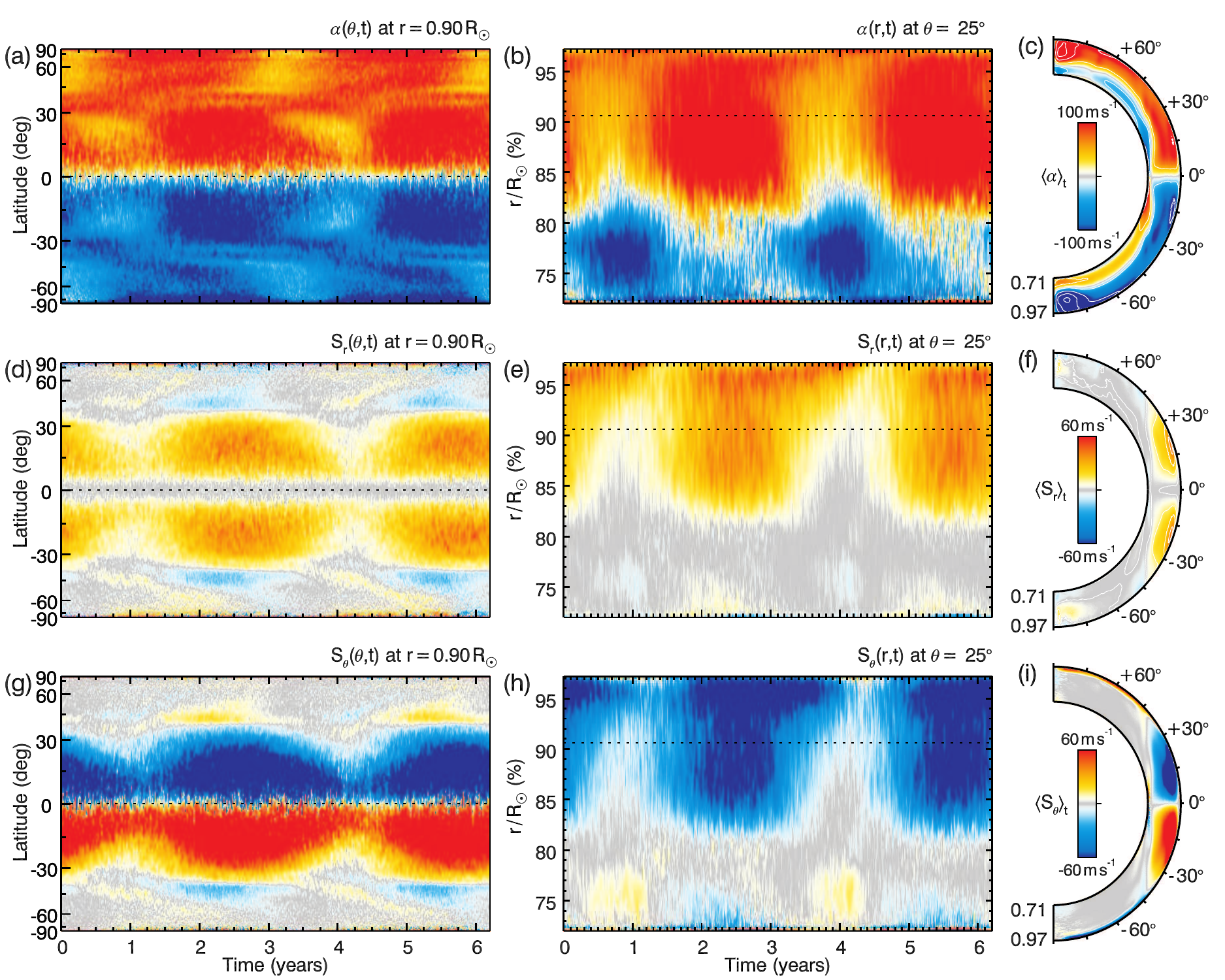} 
     \caption[Dynamo-wave interpretation of the equatorward propagation in K3S]{A mean-field
       interpretation of the K3S dynamo over the average cycle. The isotropic $\alpha$-effect
       arising from the kinetic helicity is shown (a) with latitude and time, (b) with radius and
       time, and (c) averaged over the cycle. The radial velocity $S_r$ and latitudinal velocity
       $S_{\theta}$ of the dynamo wave propagation within the context of the Parker-Yoshimura
       mechanism are exhibited in (d-i). (d) A radial cut at $\Rsun{0.90}$ through the radial
       propagation $S_r$ of the dynamo wave, outward propagation is in red and downward in blue. (e)
       A latitudinal cut through $S_r$, colors as in (d) and the dotted line indicates the depth of
       the radial cut in (d). (f) $S_r$ averaged over the cycle and shown in the meridional
       plane. Figures (g-i) are as in (d-f), but for $S_{\theta}$, with the latitudinal propagation
       is equatorward at latitudes above $30^{\circ}$ and largely poleward at lower latitudes (blue
       in the northern hemisphere and red in the south). \label{fig:d3spy}}
   \end{center}
   \vspace{-0.25truein}
\end{figure}

\section{Equatorward Dynamo Wave Propagation} \label{sec:d3spropagate}

As with ASH and EULAG, simulations in spherical segments that employ the Pencil code have also
obtained regular cyclic magnetic behavior. Some of these polarity reversing solutions exhibited
equatorward propagating magnetic features \citep{kapyla12,kapyla13}, magnetic flux ejection
\citep{warnecke12}, and 33-year magnetic polarity cycles \citep{warnecke13c}. Currently, however,
the mechanism for the equatorward propagation of the magnetic structures in those simulations
remains unclear. Perhaps the mechanism is similar to that seen here.

The equatorward propagation of magnetic features observed in K3S, as in Figures \ref{fig:d3stfe}(c)
and \ref{fig:d3sint}(a), arises through two mechanisms. The first process is the nonlinear feedback
of the Lorentz force that acts to quench the differential rotation, disrupting the convective
patterns and the shear-sustaining Reynolds stresses they possess. Since the latitudinal shear serves
to build and maintain the magnetic wreaths, the latitude of peak magnetic energy corresponds to that
of the greatest shear. So the region with the largest magnetic potential energy in the form of
latitudinal shear moves progressively closer to the equator as the Lorentz forces of the wreaths
locally weaken the shear. Such a mechanism explains the periodic modifications of the differential
rotation seen in Figure \ref{fig:d3somega}(c). However, it does not explain how this propagation is
initiated and sustained, as one might instead expect an equilibrium to be established with the
magnetic energy generation balancing the production of shear and which is further moderated by
cross-equatorially magnetic flux cancellation as the distance between the wreaths declines.

There are two possibilities for the second mechanism that promotes the equatorward propagation of
toroidal magnetic field structures. If the dynamo action in this case may be considered as a dynamo
wave, the velocity of the dynamo wave propagation is sensitive to the gradients in the angular
velocity and the kinetic helicity in the context of an $\alpha\Omega$ dynamo
\citep[e.g.,][]{parker55,yoshimura75}. This dynamo-wave velocity is given as $\mathbf{S} =
-\lambda\alpha\pht\cross\grad\Omega$, where $\lambda = r\sint$. The basic $\alpha$-effect has
$\alpha=-\tau_o \avg{\vv\boldsymbol{\cdot\omega}}/3$, and so depends on the convective overturning
time $\tau_o$ and the kinetic helicity. These quantities are shown in Figure \ref{fig:d3spy} with
time-latitude, time-radius, and time-averaged plots over the average cycle. This simple analysis
indicates that near and poleward of the edge of the low-latitude wreaths the sign of the
Parker-Yoshimura mechanism is correct to push the dynamo wave toward the equator, but the effect is
marginal elsewhere. The second possibility is that the spatial and temporal offsets between the
fluctuating EMF and the mean-shear production of toroidal field leads to a nonlinear inducement to
move equatorward. This mechanism relies on the concurrent movement of the turbulent production of
the poloidal field that continues to destroy gradients in angular velocity through the production of
toroidal magnetic through the action of the differential rotation on the renewed low-latitude
poloidal field.

\section{Conclusions} \label{sec:d3sconclude}

The ASH simulation K3S presented here is the first to self-consistently exhibit four prominent
aspects of solar magnetism: regular magnetic energy cycles during which the magnetic polarity
reverses, akin to the sunspot cycle; magnetic polarity cycles with a regular period of 6.2~years,
where the orientation of the dipole moment returns to that of the initial condition; the equatorward
migration of toroidal field structures during these cycles; and an interval of magnetic quiescence
after which the previous polarity cycle is recovered. Furthermore, this simulation may capture some
aspects of the influence of a layer of near-surface shear, with a weak negative gradient in $\avgO$
within the upper 10\% of the computation domain (3\% by solar radius). The magnetic energy cycles
with the time scale $\tau_C/2$ arise through the nonlinear interaction of the differential rotation
and the Lorentz force. The magnetic fields further quench the differential rotation by impacting the
convective angular momentum transport during the reversal. Despite the nonlinearity of the case,
there is an eligible influence of a dynamo wave in the fluctuating production of poloidal magnetic
field linked to the shear-produced toroidal field. The possible mechanisms producing the equatorward
propagation of the toroidal fields have been examined, with the nonlinear dynamo wave character of
the solution and the nonlocal interaction of the turbulent EMF and the mean-shear production of
toroidal field all playing a role. However, the dominant mechanism behind the equatorward
propagation is the nonlinear feedback of the Lorentz force on the differential rotation. This
simulation has also exhibited grand minima, similar to a Maunder minimum. Indeed, there is an
interval covering 20\% of the cycles during which the polarity does not reverse and the magnetic
energy is substantially reduced. Hence, some of the features of the dynamo active within the Sun's
interior may have been captured in this global-scale ASH simulation.

\acknowledgments{A singular thanks is due to Nicholas Featherstone for his effort in greatly
  improving the computational efficiency and scaling of the ASH code, without which this work would
  not have been possible.  The authors also thank Brad Hindman, Mark Rast, Matthias Rempel, and
  Regner Trampedach for useful conversations. The author thanks the NCAR advanced study program for
  their continuing support, as well as the support provided by NASA through the Heliophysics Theory
  Program grant NNX11AJ36G. The computations were primarily carried out on Pleiades at NASA Ames
  with SMD grants g26133 and s0943, and also used XSEDE resources for analysis. This work also
  utilized the Janus supercomputer.}


\normalsize
\begin{references}

\bibitem[{{Augustson} {et~al.}(2013){Augustson}, {Brun}, \&
  {Toomre}}]{augustson13a}
{Augustson}, K.~C., {Brun}, A.~S., \& {Toomre}, J. 2013, \apj, 777, 153

\bibitem[{{Brown} {et~al.}(2010){Brown}, {Browning}, {Brun}, {Miesch}, \&
  {Toomre}}]{brown10}
{Brown}, B.~P., {Browning}, M.~K., {Brun}, A.~S., {Miesch}, M.~S., \& {Toomre},
  J. 2010, \apj, 711, 424

\bibitem[{{Brown} {et~al.}(2011){Brown}, {Miesch}, {Browning}, {Brun}, \&
  {Toomre}}]{brown11a}
{Brown}, B.~P., {Miesch}, M.~S., {Browning}, M.~K., {Brun}, A.~S., \& {Toomre},
  J. 2011, \apj, 731, 69

\bibitem[{{Browning} {et~al.}(2006){Browning}, {Miesch}, {Brun}, \&
  {Toomre}}]{browning06}
{Browning}, M.~K., {Miesch}, M.~S., {Brun}, A.~S., \& {Toomre}, J. 2006, \apjl,
  648, L157

\bibitem[{{Brun} {et~al.}(2004){Brun}, {Miesch}, \& {Toomre}}]{brun04}
{Brun}, A.~S., {Miesch}, M.~S., \& {Toomre}, J. 2004, \apj, 614, 1073

\bibitem[{{Charbonneau}(2013)}]{charbonneau13}
{Charbonneau}, P. 2013, \nat, 493, 613

\bibitem[{{Clune} {et~al.}(1999){Clune}, {Elliott}, {Miesch}, {Toomre}, \&
  {Glatzmaier}}]{clune99}
{Clune}, T.~L., {Elliott}, J.~R., {Miesch}, M.~S., {Toomre}, J., \&
  {Glatzmaier}, G.~A. 1999, Para. Comp., 25, 361

\bibitem[{{DeRosa} {et~al.}(2012){DeRosa}, {Brun}, \& {Hoeksema}}]{derosa12}
{DeRosa}, M.~L., {Brun}, A.~S., \& {Hoeksema}, J.~T. 2012, \apj, 757, 96

\bibitem[{{Fan} {et~al.}(2013){Fan}, {Featherstone}, \& {Fang}}]{fan13}
{Fan}, Y., {Featherstone}, N., \& {Fang}, F. 2013, ArXiv e-prints

\bibitem[{{Ghizaru} {et~al.}(2010){Ghizaru}, {Charbonneau}, \&
  {Smolarkiewicz}}]{ghizaru10}
{Ghizaru}, M., {Charbonneau}, P., \& {Smolarkiewicz}, P.~K. 2010, \apjl, 715,
  L133

\bibitem[{{Gilman}(1983)}]{gilman83}
{Gilman}, P.~A. 1983, \apjs, 53, 243

\bibitem[{{Glatzmaier}(1985)}]{glatzmaier85}
{Glatzmaier}, G.~A. 1985, \apj, 291, 300

\bibitem[{{Hathaway}(2010)}]{hathaway10}
{Hathaway}, D.~H. 2010, Living Reviews in Solar Physics, 7, 1

\bibitem[{{K{\"a}pyl{\"a}} {et~al.}(2012){K{\"a}pyl{\"a}}, {Mantere}, \&
  {Brandenburg}}]{kapyla12}
{K{\"a}pyl{\"a}}, P.~J., {Mantere}, M.~J., \& {Brandenburg}, A. 2012, \apjl,
  755, L22

\bibitem[{{K{\"a}pyl{\"a}} {et~al.}(2013){K{\"a}pyl{\"a}}, {Mantere}, {Cole},
  {Warnecke}, \& {Brandenburg}}]{kapyla13}
{K{\"a}pyl{\"a}}, P.~J., {Mantere}, M.~J., {Cole}, E., {Warnecke}, J., \&
  {Brandenburg}, A. 2013, ArXiv e-prints

\bibitem[{{Malkus} \& {Proctor}(1975)}]{malkus75}
{Malkus}, W.~V.~R., \& {Proctor}, M.~R.~E. 1975, Journal of Fluid Mechanics,
  67, 417

\bibitem[{{Miesch} {et~al.}(2000){Miesch}, {Elliott}, {Toomre}, {Clune},
  {Glatzmaier}, \& {Gilman}}]{miesch00}
{Miesch}, M.~S., {Elliott}, J.~R., {Toomre}, J., {et~al.} 2000, \apj, 532, 593

\bibitem[{{Nelson} {et~al.}(2013){Nelson}, {Brown}, {Brun}, {Miesch}, \&
  {Toomre}}]{nelson13a}
{Nelson}, N.~J., {Brown}, B.~P., {Brun}, A.~S., {Miesch}, M.~S., \& {Toomre},
  J. 2013, \apj, 762, 73

\bibitem[{{Parker}(1955)}]{parker55}
{Parker}, E.~N. 1955, \apj, 122, 293

\bibitem[{{Parker}(1977)}]{parker77}
---. 1977, \araa, 15, 45

\bibitem[{{Parker}(1987)}]{parker87}
---. 1987, \apj, 312, 868

\bibitem[{{Racine} {et~al.}(2011){Racine}, {Charbonneau}, {Ghizaru}, {Bouchat},
  \& {Smolarkiewicz}}]{racine11}
{Racine}, {\'E}., {Charbonneau}, P., {Ghizaru}, M., {Bouchat}, A., \&
  {Smolarkiewicz}, P.~K. 2011, \apj, 735, 46

\bibitem[{{Rempel} {et~al.}(2009){Rempel}, {Sch{\"u}ssler}, \&
  {Kn{\"o}lker}}]{rempel09}
{Rempel}, M., {Sch{\"u}ssler}, M., \& {Kn{\"o}lker}, M. 2009, \apj, 691, 640

\bibitem[{{Steenbeck} \& {Krause}(1969)}]{steenbeck69}
{Steenbeck}, M., \& {Krause}, F. 1969, Astronomische Nachrichten, 291, 49

\bibitem[{{Tobias}(1997)}]{tobias97}
{Tobias}, S.~M. 1997, \aap, 322, 1007

\bibitem[{{Warnecke}(2013)}]{warnecke13c}
{Warnecke}, J. 2013, PhD thesis, University of Stockholm

\bibitem[{{Warnecke} {et~al.}(2012){Warnecke}, {K{\"a}pyl{\"a}}, {Mantere}, \&
  {Brandenburg}}]{warnecke12}
{Warnecke}, J., {K{\"a}pyl{\"a}}, P.~J., {Mantere}, M.~J., \& {Brandenburg}, A.
  2012, \solphys, 280, 299

\bibitem[{{Weiss} {et~al.}(1984){Weiss}, {Cattaneo}, \& {Jones}}]{weiss84}
{Weiss}, N.~O., {Cattaneo}, F., \& {Jones}, C.~A. 1984, Geophysical and
  Astrophysical Fluid Dynamics, 30, 305

\bibitem[{{Yoshimura}(1975)}]{yoshimura75}
{Yoshimura}, H. 1975, \apj, 201, 740

\end{references}
\end{document}